\documentstyle[12pt]{article}
\newcommand{\cz}{C_0}
\newcommand{\cuu}{C_{11}}
\newcommand{\cud}{C_{12}}
\newcommand{\mg}{m_{\tilde{g}}}

\newcommand{\mhs}{M^2_{H^{+}}}
\newcommand{\beq}{\begin{equation}}
\newcommand{\eeq}{\end{equation}}
\newcommand{\beqn}{\begin{eqnarray}}
\newcommand{\eeqn}{\end{eqnarray}}
\begin{document}

\thispagestyle{empty}

\begin{flushright}
{\parbox{3.5cm}{
UAB-FT-370

hep-ph/9507461

July, 1995
}}
\end{flushright}

\vspace{3cm}
\begin{center}
\begin{large}
\begin{bf}
SUPERSYMMETRIC QCD CORRECTIONS TO THE CHARGED HIGGS BOSON DECAY OF THE
TOP QUARK\\
\end{bf}
\end{large}
\vspace{1cm}
Jaume GUASCH, Ricardo A. JIM\'ENEZ, Joan SOL\`A\\

\vspace{0.25cm}
Grup de F\'{\i}sica Te\`orica\\
and\\
Institut de F\'\i sica d'Altes Energies\\
\vspace{0.25cm}
Universitat Aut\`onoma de Barcelona\\
08193 Bellaterra (Barcelona), Catalonia, Spain\\
\end{center}
\vspace{0.3cm}
\hyphenation{super-symme-tric}
\hyphenation{com-pe-ti-ti-ve}
\begin{center}
{\bf ABSTRACT}
\end{center}
\begin{quotation}
\noindent
The one-loop supersymmetric QCD quantum effects
on the width of the unconventional top quark decay mode
$t\rightarrow H^{+}\, b$ are evaluated within the MSSM.
The study of this process is useful to
hint at the supersymmetric nature of the charged Higgs
emerging from that decay.
Remarkably enough, recent calculations of
supersymmetric corrections to $Z$-boson observables have
shown that the particular conditions by which the decay
$t\rightarrow H^{+}\, b$ becomes competitive with the
standard decay $t\rightarrow W^{+}\,b$ have a chance to
be realized in nature. This further motivates us to focus our attention
on the dynamics of $t\rightarrow H^{+}\, b$ as an excellent laboratory
to unravel Supersymmetry at the quantum level in
future experiments at Tevatron and at LHC.
\end{quotation}

\baselineskip=6.5mm  

\newpage

The recent discovery of the top quark at Tevatron\,\cite{CDF,D0} is
reckoned to be the latest major
event in the world of elementary particle physics. The weighted average
of the $CDF$ and $D0$ measurements yields a mass,
$m_t=180\pm 12\,GeV$, which is in good agreement with the Standard Model (SM)
global analyses of the electroweak precision data. It can also be used to
sharpen
the prediction of the SM Higgs mass, $M_H$, in a range centered
at $\sim 100\,GeV$ or below\,\cite{EFL}. Far from being a final confirmation
of the SM, the finding of the top quark raises many questions on the nature
of the spontaneous symmetry-breaking mechanism (SSB) that go beyond
the SM, in particular whether the SSB is caused by fundamental scalars or by
some dynamical mechanism postulating a new species of strongly-interacting
fermions (such as in technicolour like models\,\cite{TC}), or perhaps involving
the condensation of the top quark itself (such as in topcolour models
\,\cite{Hill}).
In this paper we shall adhere to the extensions of the SM associated with
elementary scalars, specifically to the Minimal Supersymmetric Standard
Model MSSM\,\cite{MSSM}. The global fit analyses of
precision data within the MSSM chart an increasing trend of compatibility
with the MSSM\,\,\cite{ChankPok}.
They lead to a top quark mass, $m_t=165\pm 10\,GeV$, consistent
with the experimental measurements.
Moreover, the richer structure of the top and Higgs sector of the MSSM gives
rise to a very stimulating top/stop-Higgs/higgsino dynamics.
The archetype example of it could be the non-standard decay
of the top quark into a charged Higgs: $t\rightarrow H^+\,b$. Although there
are in principle many new exotic decays of the top quark in the MSSM,
perhaps the latter mode is (if open) the closest one to the canonical
mode $t\rightarrow W^+\,b$ in the SM and therefore the less difficult to handle
from the experimental point of view.

Apart from its obvious interest on its own, a further motivation\,\cite{JSP}
to consider the analysis of the charged Higgs decay of the top quark stems
from the recent results of $Z$ physics, particularly
of the observed anomaly in the ratios $R_b$, $R_c$ as well as in the value of
$\alpha_s(M_Z)$ \,\cite{LEPWG}. These anomalies are
discrepancies (at the $2-3\,\sigma$ level) between the measured
values of these observables and the corresponding predictions of the SM.
As shown by early calculations of supersymmetric (SUSY) radiative corrections
to $R_b$\,\cite{Rb1}, the theoretical prediction of this observable
can be made in better agreement with
experiment. More recently, there has been a flood of renewed
interest in this subject (see e.g.
\,\cite{Rb2}-\cite{Rb3}) and it has been possible to establish in a more
precise
way the particular conditions by which the MSSM is able to cure,
or at least to alleviate, the ``$R_b$ crisis''. The corresponding MSSM
analysis of the ratio $R_c$ (in correlation with the ratio $R_b$) was first
presented in Ref.\cite{GS1}. The upshot is that both the $R_b$ and $R_c$
``crises'' can be solved within the MSSM  provided that
$\tan\beta$ is large enough
\footnote{A possible solution also exists for $\tan\beta<1$,
but it is not favoured by model building\,\cite{Giudice} and it is not relevant
to
the present analysis.} and there exists a
supersymmetric pseudoscalar Higgs as well as some superpartner all of
them in the $50\,GeV$ range --in agreement with the most
recent global fit analyses\,\cite{ChankPok}.
In these conditions it turns out that one can simultaneously
provide a supersymmetric explanation\,\cite{GS2}
of the longstanding mismatch between the low-energy and high-energy
determinations of $\alpha_s(M_Z)$\,\cite{Shifman}. All in all, these results
bring in an independent incentive from the high precision world of $Z$ physics
to choose our SUSY parameters in the region of
large $\tan\beta$ and moderate charged Higgs mass. Indeed,
from the well-known Higgs mass relations in the MSSM\,\cite{Hunter} and
assuming that a light CP-odd (``pseudoscalar'') Higgs mass $m_{A^0}$
exists in the $50\,GeV$ ballpark\footnote{It is noteworthy that
at high $\tan\beta$ the approximate
phenomenological lower limit
$m_{h^0}\stackrel{\scriptstyle >}{{ }_{\sim}} 50\,GeV$ on the
light CP-even Higgs mass of the MSSM translates into
$m_{A^0}\stackrel{\scriptstyle >}{{ }_{\sim}} 50\,GeV$ for the CP-odd mass.
Recent global fit analyses also favour a light Higgs mass in the SM and a
light CP-even Higgs mass in the MSSM\,\cite{EFL}. },
it follows that there must be a charged Higgs companion of
$M_{H^{\pm}}\simeq 100\, GeV$. In these circumstances
the decay  $t\rightarrow H^{+}\, b$ becomes
competitive with the SM decay $t\rightarrow W^{+}\, b$
and it should be possible to identify it by tagging violations of
lepton universality caused by the presence
of an excess of
final state $\tau$-leptons associated to the subsequent Higgs decay --as
commented at the end of Ref.\cite{GS1}.

In view of the potential interest of the decay mode $t\rightarrow H^{+}\, b$,
one would naturally like to address the computation of the strong virtual
corrections to its partial width. Of these, the conventional QCD corrections
have already been considered in detail in Ref.\cite{CD} and they turn out to be
sizeable and negative (of order $-10\%$). Although they are blind to the
nature of the underlying Higgs model, they need to be subtracted from the
experimentally measured number in order to be be able to
probe the existence of new sources of quantum effects
beyond the SM. These effects may ultimately reveal whether the charged Higgs
emerging from that decay is supersymmetric or not.

In this paper we compute the strong SUSY radiative corrections to the
partial width by paying special attention to the aforementioned
privileged region of the MSSM parameter space. The analysis
of the larger and more complex
body of SUSY electroweak corrections, namely the corrections mediated by
squarks, sleptons, chargino-neutralinos and the Higgs bosons themselves, will
be
presented elsewhere\,\cite{CGGJS}\footnote{The study of the corresponding
supersymmetric quantum corrections to
the canonical decay $t\rightarrow W^+\,b$ has been presented by some of
the authors in Refs.\cite{GJSH} and \cite{DHJJS} (See also
Ref.\cite{YangLi}).}.
To compute the one-loop QCD corrections to
 $\Gamma_t\equiv\Gamma (t\rightarrow H^{+}\, b)$ in the MSSM, we shall adopt
the on-shell renormalization scheme where the fine structure constant,
$\alpha$, and the masses of the gauge bosons, fermions and scalars are
the renormalized parameters: $(\alpha, M_W, M_Z, M_H, m_f, M_{SUSY},...)$
\,\cite{BSH}.
The interaction Lagrangian describing the $t\,b\,H^{\pm}$-vertex
in the MSSM reads as follows:
\beq
{\cal L}_{Hbt}={g\,V_{tb}\over\sqrt{2}M_W}\,H^-\,\bar{b}\,
[m_t\cot\beta\,P_R + m_b\tan\beta\,P_L]\,t+{\rm h.c.}\,,
\label{eq:LtbH}
\eeq
where $P_{L,R}=1/2(1\mp\gamma_5)$ are the chiral projector operators,
 $\tan\beta$ is the ratio between the vacuum expectation values of the
two Higgs doublets of the MSSM\,\cite{MSSM} and
$V_{tb}$ is the corresponding Kobayashi-Maskawa matrix element--hereafter
we set $V_{tb}=1$  ($V_{tb}=0.999$ within
$\pm 0.1\%$, from unitarity of the KM-matrix {\it and} assuming $3$ quark
families).

There are no oblique strong supersymmetric corrections at 1-loop
order. The non-oblique vertex corrections originating from gluinos and squarks
(stop and sbottom species) are depicted in Fig.1.
The SUSY-QCD interaction Lagrangian relevant to our calculation is given,
in four-component notation, by
\begin{equation}
{\cal L}= -{g_s\over\sqrt{2}}\,\left[\tilde{q}^{i *}_{L}\,(\lambda_r)_{ij}\,
\bar{\tilde{g^r}}\,P_L\,q^j-\bar{q}^i(\lambda_r)_{ij}\,
P_L\,{\tilde{g^r}}\,\tilde{q}^{j}_{R}\right]+{\rm h.c.}\,,
\end{equation}
where $\tilde{g}^r (r=1,2,...,8)$ are the Majorana gluino fields,
$(\lambda_r)_{ij} (i,j=1,2,3) $ are the Gell-Mann matrices, and
$\tilde{q'}_a=\{\tilde{q}_L,\tilde{q}_R\}$ are the weak-eigenstate squarks
associated to the two chiral components  $P_{L,R}\,q$;
they are related to
the corresponding mass-eigenstates $\tilde{q}_a=\{\tilde{q}_1,\tilde{q}_2\}$
by a rotation $2\times 2$ matrix (we neglect intergenerational mixing):
\begin{eqnarray}
\tilde{q'}_a&=&\sum_{b} R_{ab}^{(q)}\tilde{q}_b,\nonumber\\
R^{(q)}& =&\left(\begin{array}{cc}
\cos{\theta_q}  &  \sin{\theta_q} \\
-\sin{\theta_q} & \cos{\theta_q}
\end{array} \right)\;\;\;\;\;\;
(q=t, b)\,.
\label{eq:rotation}
\end{eqnarray}
These rotation matrices diagonalize the corresponding stop and sbottom mass
matrices:
\begin{equation}
{\cal M}_{\tilde{t}}^2 =\left(\begin{array}{cc}
M_{\tilde{t}_L}^2+m_t^2+\cos{2\beta}({1\over 2}-
{2\over 3}\,s_W^2)\,M_Z^2
 &  m_t\, M_{LR}^t\\
m_t\, M_{LR}^t &
M_{\tilde{t}_R}^2+m_t^2+{2\over 3}\,\cos{2\beta}\,s_W^2\,M_Z^2\,.
\end{array} \right)\,,
\label{eq:stopmatrix}
\end{equation}
\begin{equation}
{\cal M}_{\tilde{b}}^2 =\left(\begin{array}{cc}
M_{\tilde{b}_L}^2+m_b^2+\cos{2\beta}(-{1\over 2}+
{1\over 3}\,s_W^2)\,M_Z^2
 &  m_b\, M_{LR}^b\\
m_b\, M_{LR}^b &
M_{\tilde{b}_R}^2+m_b^2-{1\over 3}\,\cos{2\beta}\,s_W^2\,M_Z^2\,,
\end{array} \right)\,,
\label{eq:sbottommatrix}
\end{equation}
with
\beq
M_{LR}^t=A_t-\mu\cot\beta\,, \ \ \ \ M_{LR}^b=A_b-\mu\tan\beta\,,
\eeq
$\mu$ being the SUSY Higgs mass parameter in the superpotential\footnote{Its
sign is relevant in the numerical analysis. We fix it as in
eq.(3) of Ref.\cite{GJSH}.}.
The $A_{t,b}$ are the trilinear soft SUSY-breaking parameters and the
$M_{{\tilde{q}}_{L,R}}$ are soft SUSY-breaking masses\,\cite{MSSM}.
By $SU(2)_L$-gauge invariance we must have $M_{\tilde{t}_L}=M_{\tilde{b}_L}$,
whereas $M_{{\tilde{t}}_R}$, $M_{{\tilde{b}}_R}$ are in general independent
parameters.
Finally, we also need the interaction Lagrangian involving the charged Higgs
and the stop and sbottom squarks
\beq
{\cal L}_{H\tilde{b}\tilde{t}}=-{g\over \sqrt{2}\,M_W}\,H^-\,\left(
g_{LL}\,\tilde{b}_L^*\,\tilde{t}_L+g_{RR}\,\tilde{b}_R^*\,\tilde{t}_R
+g_{LR}\,\tilde{b}_R^*\,\tilde{t}_L+g_{RL}\,\tilde{b}_L^*\,\tilde{t}_R\right)\
 +{\rm h.c.}\,,
\label{eq:Htildebt}
\eeq
where
\beqn
g_{LL}&=&M_W^2\sin 2\beta -(m_t^2\cot\beta+m_b^2\tan\beta)\,,\nonumber\\
g_{RR}&=&-m_tm_b(\tan\beta+\cot\beta)\,,\nonumber\\
g_{LR}&=&-m_b(\mu+A_b\tan\beta)\,,\nonumber\\
g_{RL}&=&-m_t(\mu+A_t\cot\beta)\,.
\eeqn
The one-loop renormalized vertex, $\Lambda$, is derived from the
renormalized Lagrangian
plus counterterms, ${\cal L}\rightarrow {\cal L}+\delta {\cal L}$, following
the standard procedure\,\cite{BSH}. It
can be parametrized in terms of two form factors
$F_L$, $F_R$ and the corresponding mass and wave-function renormalization
constants $\delta m_f$,
$\delta Z_{L,R}^f$ associated to the external quarks, viz.
\beq
\Lambda = {i\,g\over\sqrt{2}\,M_W}
\,\left[m_t\,\cot\beta\,(1+\Lambda_R)\,P_R
 + m_b\,\tan\beta\,(1+\Lambda_L)\,P_L\right]\,,
\label{eq:AtbH}
\eeq
with
\beqn
\Lambda_R & = & F_R+{\delta m_t\over m_t}
+\frac{1}{2}\,\delta Z_L^b+\frac{1}{2}\,\delta Z_R^t
 \,,\nonumber\\
\Lambda_L &=& F_L+{\delta m_b\over m_b}
+\frac{1}{2}\,\delta Z_L^t+\frac{1}{2}\,\delta Z_R^b\,.
\eeqn
In the on-shell scheme we have\footnote{Our sign conventions for the
self-energy functions are those of Ref.\cite{GJSH}.}
\beq
{\delta m_q\over m_q}=-\left[{\Sigma^q_L(m_q^2)+\Sigma^q_R(m_q^2)\over 2}
+\Sigma^q_S(m_q^2)\right]
\label{eq:deltamf}
\eeq
and
\beqn
\delta Z_{L,R}^q &=& \Sigma^q_L(m_f^2)+m_q^2[\Sigma^{q\,\prime}_L
(m_q^2)+\Sigma^{q\,\prime}_R(m_q^2)
+2\Sigma^{q\,\prime}_S(m_q^2)]\,.
\label{eq:DSRC}
\eeqn
In these equations we have decomposed the (real part of the) quark
self-energy according to
\begin{equation}
\Sigma^f(p)=\Sigma^f_L(p^2)\not{p}\,P_L+\Sigma^f_R(p^2)\not{p}\,P_R
+m_f\,\Sigma^f_S(p^2)\,,
\label{eq:Sigma}
\end{equation}
and used the notation $\Sigma'(p)\equiv \partial\Sigma(p)/\partial p^2$.

{}From the renormalized amplitude (\ref{eq:AtbH}), the width
$\Gamma=\Gamma (t\rightarrow H^+\,b)$
including the one-loop SUSY-QCD corrections is the following:
\beq
\Gamma = \Gamma_0 \left\{1+\frac{N_L}{N}\,[2\,Re(\Lambda_L)]+
\frac{N_R}{N}\,[2\,Re(\Lambda_R)]+\frac{N_{LR}}{N}\,
[2\,Re(\Lambda_L+\Lambda_R)]\right\}\,,
\label{eq:1Lwidth}
\eeq
where the corresponding lowest-order result is
\beq
\Gamma_0=\left({G_F\over 8\pi\sqrt{2}}\right){N\over m_t}\,
\lambda^{1/2} (1, {m_b^2\over m_t^2},{\mhs\over m_t^2})\,.
\label{tree}
\eeq
We have defined
\begin{equation}
\lambda^{1/2} (1, x^2, y^2)\equiv\sqrt{[1-(x+y)^2][1-(x-y)^2]}
\end{equation}
and
\beqn
N &=& (m_t^2+m_b^2-\mhs)\,(m_t^2\cot^2\beta+m_b^2\tan^2\beta)
+4m_t^2m_b^2\,,\nonumber\\
N_L & = & (m_t^2+m_b^2-\mhs)\,m_b^2\tan^2\beta\,,\nonumber\\
N_R & = & (m_t^2+m_b^2-\mhs)\,m_t^2\cot^2\beta\,,\nonumber\\
N_{LR} & = & 2m_t^2m_b^2\,.
\eeqn
Notice that, in contradistinction to electroweak one-loop
calculations\,\cite{GJSH},
an additional correction term $\Delta r$
does not appear in (\ref{eq:1Lwidth}) due to the
absence of one-loop SUSY-QCD corrections in $\mu$-decay.

The explicit contribution to the form factors $F_L$,$F_R$
from the vertex diagram of Fig.1 is given by
\beqn
F_L&=&8\pi\alpha _s\,i C_F\frac{G_{ab}}{m_b \tan\beta}
[R^{(t)*}_{2 b}R^{(b)}_{2 a}(\cuu-\cud)m_t+
R^{(t)*}_{1 b}R^{(b)}_{1 a}\cud m_b+
R^{(t)*}_{1 b}R^{(b)}_{2 a}\cz \mg] \,,\nonumber\\
F_R&=&8\pi\alpha _s\,i C_F\frac{G_{ab}}{m_t \cot\beta}
[R^{(t)*}_{1 b}R^{(b)}_{1 a}(\cuu-\cud)m_t+
R^{(t)*}_{2 b}R^{(b)}_{2 a}\cud m_b+
R^{(t)*}_{2 b}R^{(b)}_{1 a}\cz \mg]\,.\nonumber\\
\label{eq:FLFR}
\eeqn
Here
$C_F=(N_C^2-1)/2N_C=4/3$ is a colour factor.
We have furthermore defined:
\beq
G_{ab}=R^{(t)}_{1 b}R^{(b)*}_{1 a} g_{LL}+
R^{(t)}_{2 b}R^{(b)*}_{2 a}g_{RR}+
R^{(t)}_{1 b}R^{(b)*}_{2 a}g_{LR}+
R^{(t)}_{2 b}R^{(b)*}_{1 a} g_{RL}\,.
\eeq
The three-point function notation is as in Ref.\cite{GJSH,Axelrod},
with the following arguments:
\beq
C=C(p,p^{\prime},m_{\tilde{g}},m_{\tilde{t}_b},m_{\tilde{b}_a})\,.
\eeq
As for the self-energies,
\beqn
\Sigma^q_L(p^2)&=&8\pi\alpha_s\,i C_F
\,|R^{(q)}_{1 a}|^2(B_0-B_1)\,,\nonumber\\
\Sigma^q_R(p^2)&=&8\pi\alpha_s\,i C_F
\,|R^{(q)}_{2 a}|^2(B_0-B_1)\,,\nonumber\\
\Sigma^q_S(p^2)&=&-\,8\pi\alpha_s\,i C_F
\frac{m_{\tilde{g}}}{m_q}\, {\rm Re}(\,R^{(q)}_{1 a}\,R^{(q)*}_{2 a}) B_0\,,
\label{eq:selfs}
\eeqn
where the two-point functions --defined also as in Ref.\cite{GJSH}-- have the
following arguments:
\beq
B=B(p^2,m^2_{\tilde{q}_a},m^2_{\tilde{g}})\,.
\eeq
(A summation over squark indices is understood in eqs.(\ref{eq:FLFR}) and
(\ref{eq:selfs}).)
It is easy to convince oneself that the form factors (\ref{eq:FLFR}) are
to be $UV$-finite in SUSY-QCD, as indeed they are.
The remaining contributions to the
renormalized amplitude (\ref{eq:AtbH}) are immediately seen to cancel
$UV$-divergences each other out.

The numerical analysis of the strong supersymmetric
corrections to $\Gamma (t\rightarrow H^+\,b)$
is exhibited in Figs.2-5. We present the results both in terms of the corrected
$\Gamma$ and in terms of the
relative correction with respect to the tree-level width, i.e.
\begin{equation}
\delta_{\tilde{g}}={\Gamma-\Gamma_0\over \Gamma_0}\,.
\label{eq:deltag}
\end{equation}
The free parameters
at our disposal lie in the mass matrices
(\ref{eq:stopmatrix})-(\ref{eq:sbottommatrix}).
Among the sfermion mass matrices, the stop mass matrix is the only one
where a non-diagonal structure caused by a sizeable mixing term is most
likely to arise.
For $M_{LR}^t=0$ this matrix is trivial, but since $m_t$ is
large a nonvanishing $M_{LR}^t$ naturally leads to
a light mass eigenvalue, denoted by $m_{\tilde{t}_1}$, whereas the other
eigenvalue, $m_{\tilde{t}_2}$, can be much heavier.
As a matter of fact a light stop with a mass
$m_{\tilde{t}_1}={\cal O}(M_Z/2)$ is still
phenomenologically allowed\,\cite{Baer}.
In contrast, the off-diagonal element of the sbottom mass matrix
(\ref{eq:sbottommatrix}), being proportional to $m_b\simeq 4.5\,GeV$,
is expected to be small (unless $M_{LR}^b$ is very large).
We shall treat the sbottom mass matrix in the
simplest possible way compatible
with the phenomenological bounds on
squark masses\,\cite{Baer}-- only scaped perhaps by the lightest stop.
For definitiveness, unless stated otherwise,
we shall assume that $M_{LR}^b=0$ (equivalently $A_b=\mu\tan\beta$) and that
the two mass eigenvalues are equal
($m_{\tilde{b}_1}=m_{\tilde{b}_2}\equiv m_{\tilde{b}}$) and constrained to
satisfy $m_{\tilde{b}}\geq 150\,GeV$.
As for $A_t$, it will be treated either as an input parameter or it will be
fixed once we are given $M_{LR}^t$, $\mu$ and $\tan\beta$. We remind
that $M_{LR}^t$ is expected to preserve the inequality
\begin{equation}
M_{LR}^t\leq 3\,m_{\tilde{b}_L}\,,
\label{eq:MLR}
\end{equation}
which roughly corresponds to a necessary, though not sufficient, condition to
 avoid colour-breaking vacua
\,\cite{Frere}.
In the conditions described above, once
 $m_{\tilde{b}}$, $\mu$ and $\tan\beta$
are fixed, the stop mass matrix depends on only $2$ parameters, e.g.
$(A_t,M_{\tilde{t}_R})$, $(M_{LR}^t,m_{\tilde{t}_1})$, etc.
For the strong coupling constant we used the value
\begin{equation}
\alpha_s={g_s^2\over 4\pi}=0.11
\end{equation}
which remains essentially constant within the CDF-D0 ranges mentioned above.
Whenever $m_t$ needs to be fixed,
we take the central value $m_t=180\,GeV$.

A crucial parameter to be explored in our analysis is $\tan\beta$.
In Fig.2a we plot the SUSY-QCD corrected
$\Gamma=\Gamma(t\rightarrow H^+\,b)$, eq.(\ref{eq:1Lwidth}),
versus $\tan\beta$ for $\mu=+100\,GeV$ and $\mu=-100\,GeV$, and for
given values of the other parameters. We see that $\Gamma$ is very
sensitive to $\tan\beta$ and that, barring the narrow interval $\tan\beta\leq
1$,
the process $t\rightarrow H^+\,b$ becomes steadily competitive with
the standard process, $t\rightarrow W^+\,b$, in the large $\tan\beta$ region,
i.e.
when $\tan\beta$ is of the order of $30-40\simeq m_t/m_b$.
As mentioned in the beginning, this is precisely
the range singled out by the $Z$ boson observables\,\cite{GS1}.
We also see from Fig.2a that, for
$\tan\beta \stackrel{\scriptstyle >}{{ }_{\sim}} 30$, $\Gamma$
starts to deviate from $\Gamma_0$ quite manifestly. Therefore we shall choose
$\tan\beta=30$ as a representative value in the other plots.

Highly remarkable is also the incidence of the parameter $\mu$
both of its value and of its sign.
In fact, the sign of  $\delta_{\tilde{g}}$ happens to be opposite
to the sign of $\mu$ and the respective corrections for $\mu$ and for $-\mu$
take on approximately the same
absolute value. In Fig.2b we deliver the correction itself,
$\delta_{\tilde{g}}$,
for $\mu = -100\,GeV$ and for different values of the squark and gluino
masses.
The sign dependence of $\delta_{\tilde{g}}$ suggests that
two extreme scenarios could take place with the SUSY-QCD
corrections to $\Gamma (t\rightarrow H^+\,b)$: namely,
they could either significantly enhance the, negative, conventional
QCD corrections\,\cite{CD},
or on the contrary they could counterbalance them and even result in
opposite sign.

In Figs.3a-3b we display $\delta_{\tilde{g}}$ as a function of $m_{\tilde{g}}$
and $m_t$, respectively, for three values of the sbottom masses.
 We learn from Fig.3a that
light gluinos of ${\cal O}(1)\,GeV$\,\cite{Clavelli} yield, contrary to naive
expectations, a rather small correction as compared to heavy gluinos
of ${\cal O}(100)\,GeV$.
It should be clear that these corrections
do eventually decouple --as we have checked-- for larger and larger
gluino masses. Notwithstanding, the decoupling rate of the gluinos is
particularly noticeable, for it happens to be so slow (Fig.3a) that it
fakes for a while a non-decoupling behaviour.
This trait is caused by the presence of a long
sustained local maximum (or minimum, depending on the sign of $\mu$)
spreading over a wide range of heavy gluino masses
centered at $\sim 300\,GeV$. For this reason, heavy gluinos
are in the present instance preferred to light gluinos.
In Fig.4, we test the sensitivity of $\delta{\tilde{g}}$
to $M_{H^{\pm}}$; it turns out to be very small, except
near the uninteresting vicinity of the
phase space border where $\Gamma_0$ is about to vanish.

In all previous figures we have fixed $A_t=0$,
 $\theta_t=\pi/4$ and $m_{\tilde{b}}\leq 150\,GeV$,
so that the lightest stop mass was always $m_{\tilde{t}_1}>190\,GeV$. In Fig.5
we relax these conditions and plot contour isolines
$\delta_{\tilde{g}}={\rm const.}$
in the $(M_{LR}^t, m_{\tilde{t}_1})$-plane, where $A_t$ and
$\theta_t$ are variable. In particular, we approach the region of
the lightest possible stop masses compatible with the strict LEP
phenomenological bound. As expected, the corrections are larger the smaller is
$m_{\tilde{t}_1}$.
Finally, let us mention that a non-vanishing mixing in the sbottom mass matrix
does not alter at all the typical size of the corrections obtained here.
It was only to avoid much cluttering of free parameters that we have
treated so far that matrix in the most simple-minded form
($M_{LR}^b=0$ and $m_{\tilde{b}_1}=m_{\tilde{b}_2}$).
For example, if the mass-eigenvalues are
$m_{\tilde{b}_{1,2}}=200,250\,GeV$ and $\mu=-100\,GeV$, $\tan\beta=30$,
then one has $A_b=-500\,GeV$ and the corresponding correction reads
$\delta_{\tilde{g}}=+30\%$. In the same conditions, but
choosing $A_b=0$,  $\mu$ is determined to be $-83\,GeV$ and
$\delta_{\tilde{g}}=+25\%$, etc.

Some words on previous work are in order.
We chart significant differences in our complete analysis as compared to
preliminary calculations in the literature.
In Ref.\cite{LiYangHu} a first study
of the SUSY-QCD corrections to $t\rightarrow H^+\,b$ was presented, but
they neglect the bottom quark Yukawa coupling and as a consequence they are
incorrectly  sensitive to the high $\tan\beta$ effects. Indeed, the form factor
$F_L$ which is associated to that coupling is by far
the dominant piece of our
numerical analysis in the large $\tan\beta$ region.
Furthermore, the impact
from mixing effects and the incidence of the various parameter dependences
were completely missed and only the simplest
situation, characterized by degenerate masses,
was considered\footnote{Notice that
the assumption of stop masses equal
to sbottom masses is incompatible with
eqs.(\ref{eq:stopmatrix})-(\ref{eq:sbottommatrix}).}.
The study of Ref.\cite{Koenig} also neglects the bottom quark Yukawa coupling.
Thus the purported large effects claimed in the large $\tan\beta$ region
are not correctly justified.
Moreover, in the framework of these two references,
the lowest-order width is fully proportional to $\cot\beta$; thus, in such a
context, finding quantum effects increasing with $\tan\beta$ is rather useless
since they result in corrections to an uninteresting,
vanishingly small, tree-level width.
Finally, the latter reference also neglects the wave-function
and mass renormalization contributions
which, numerically, are of the same order of magnitude as the vertex
contributions
in the intermediate and small $\tan\beta$ region.
However, where it has been possible to force an overlapping, we have found
numerical coincidence.

In summary,
the SUSY-QCD contributions to the partial width of $t\rightarrow H^+\,b$
could be quite large (several $10\%$); and what is more,
these corrections apply to a decay mode which, in the region of the MSSM
parameter space prompted by the high precision $Z$
boson observables\,\cite{GS1,GS2},
has an appreciable branching ratio as compared to the standard decay
$t\rightarrow W^+\,b$. Furthermore, we have found that the impact of
the gluino corrections on $t\rightarrow H^+\,b$ could occur in two
opposite ways, either by reinforcing the conventional QCD corrections or
on the contrary by severely
cancelling them out--perhaps even to the extend of reversing their sign!.
Most remarkable, the potential size  of these effects
stems not only from the strong interaction character of the
SUSY-QCD corrections, but also from the high sensitivity of
$t\rightarrow H^+\,b$ to the (weak-interaction) SSB
parameter $\tan\beta$. At the end of the day we
must conclude that $t\rightarrow H^+\,b$ could reveal itself as the ideal
environment where to study the nature of the SSB mechanism. It could even
be the right place where to target our long and unsuccessful
search for large, {\it and} slowly decoupling,
quantum supersymmetric effects. In this respect it should not be
understated the fact that the typical size of our corrections is maintained
even for sparticle masses well above the LEP $200$ discovery range.
Theses features are in stark contrast to the standard decay
of the top quark, $t\rightarrow W^+\,b$, whose SUSY-QCD corrections
are largely insensitive to $\tan\beta$\,\cite{DHJJS}. Fortunately,
the next generation of experiments at Tevatron and the future high precision
experiments at LHC may well acquire the ability to test the kind of effects
considered here (Cf. Refs.\,\cite{Atlas,Raychau,CPYuan}).
Thus, in favorable circumstances, we should be able to disentangle
the potential supersymmetric nature of the charged Higgs decay of the top
quark out of a measurement of the top quark width\footnote{Or related
observables,
such as e.g. the --double and single-- top quark production cross-sections, or
the differential distributions of the lepton final states in given
exclusive decay channels, etc.}
at a modest precision of $\sim 5-10\%$.

\newpage

{\bf Acknowledgements}:
The work of RJ and JS has  been partially supported by CICYT
under project No. AEN93-0474. The work of JG has also been financed by a
grant of the Comissionat per a Universitats i Recerca, Generalitat de
Catalunya.


\begin{center}
\begin{Large}
{\bf Figure Captions}
\end{Large}
\end{center}
\begin{itemize}
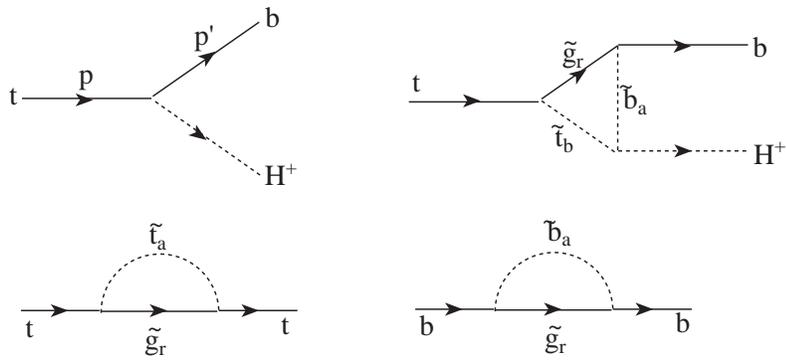
\item{\bf Fig.1} SUSY-QCD Feynman diagrams, up to one-loop order,
correcting the partial width $\Gamma (t\rightarrow H^{+}\,b)$.
Each one-loop diagram is summed over the mass-eigenstates of the
stop, sbottom squarks
($\tilde{b}_a, \tilde{t}_b\,; a,b=1,2$) and gluinos
 $\tilde{g}_r\,; r=1,2,...,8$.

\item{\bf Fig.2} (a) SUSY-QCD corrected $\Gamma(t\rightarrow H^+\,b)$ as
a function of $\tan\beta$, for two opposite values of $\mu$, compared
to the corresponding tree-level width, $\Gamma_0$. The framed set
of inputs is common to Figs. 2,3.
The horizontal (dashed) line marks $\Gamma_0(t\rightarrow W^+\,b)=1.71\,GeV$
--the tree-level width of the standard process ($m_t=180\,GeV$);
its SUSY-QCD corrections are
generally small and essentially independent of $\tan\beta$\,\cite{DHJJS};
(b) The relative correction  $\delta_{\tilde{g}}$ as a function of
$\tan\beta$ for $\mu=-100\,GeV$ and three
values of $m_{\tilde{b}}=m_{\tilde{g}}$.

\item{\bf Fig.3} Dependence of $\delta_{\tilde{g}}$ upon (a) $m_{\tilde{g}}$,
including the light gluino region, for the same squark masses
and $\mu$ as in Fig.2b;
(b) $\delta_{\tilde{g}}$ as a function of $m_t$ (within the
$CDF$-$D0$ limits) and remaining parameters as in (a).

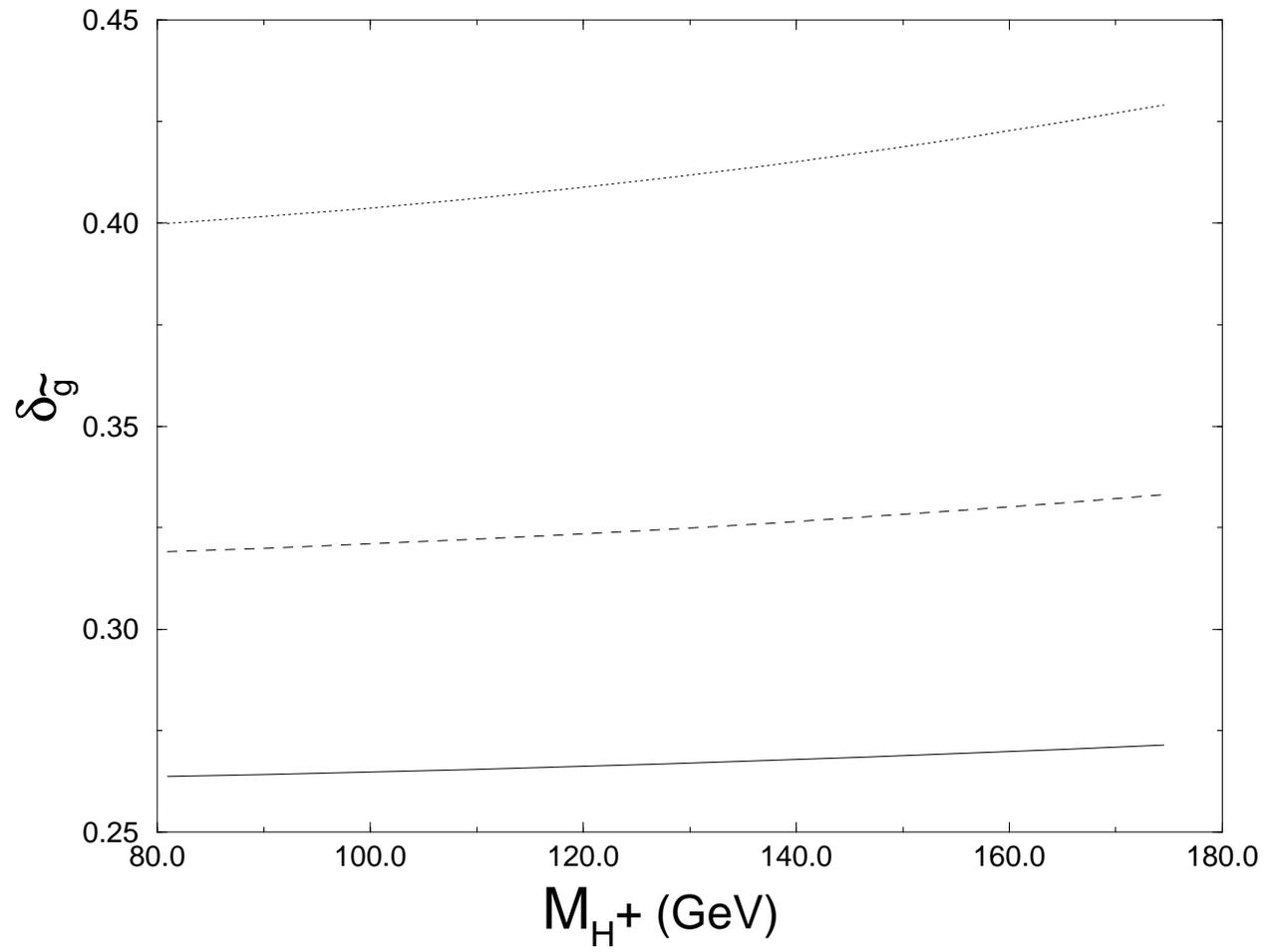
\item{\bf Fig.4}\ \ $\delta_{\tilde{g}}$ as a function of $M_{H^{\pm}}$.
Rest of inputs as in Fig.3.

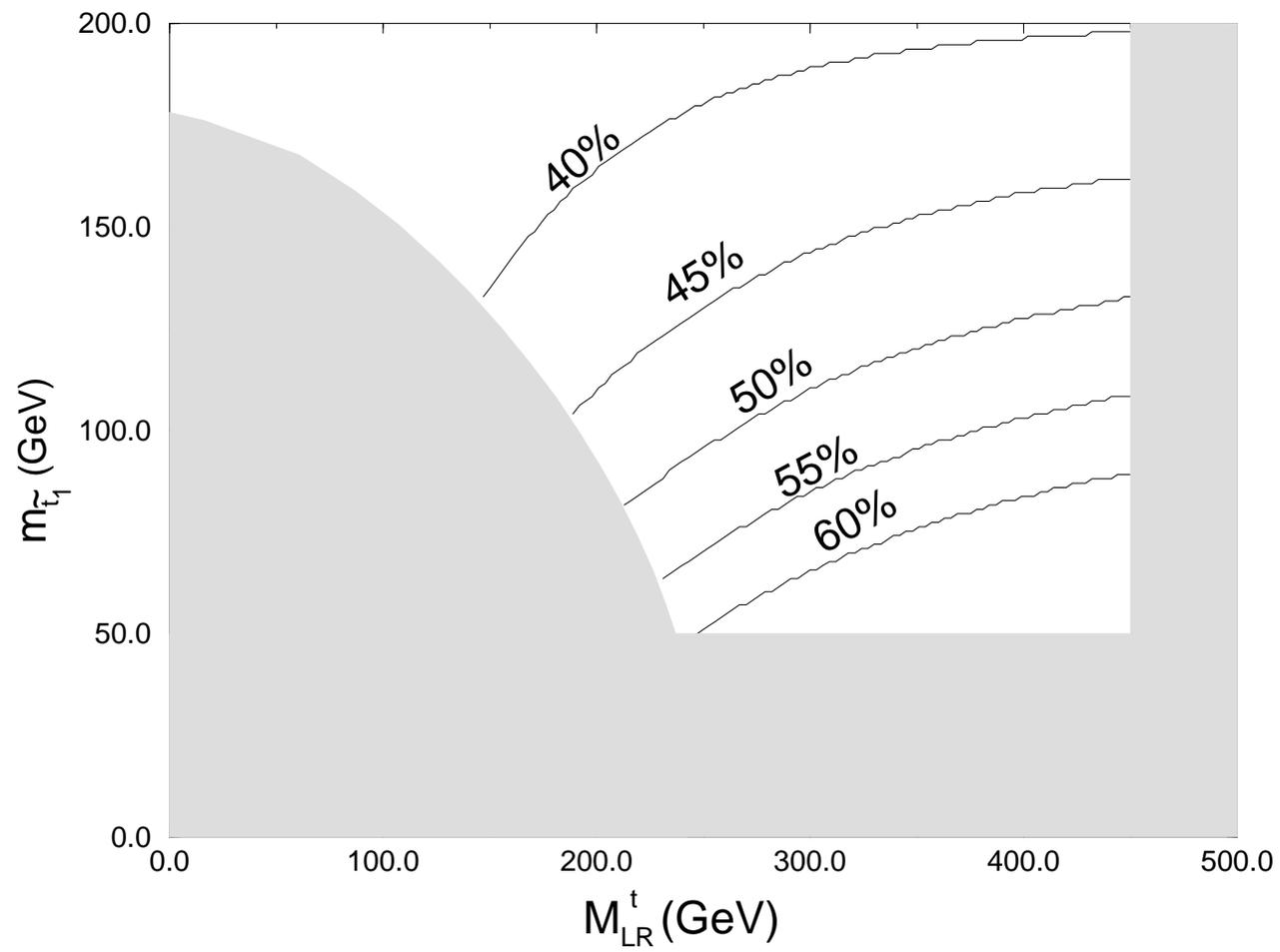
\item{\bf Fig.5} Contour plots of $\delta_{\tilde{g}}$ in the
$(M_{LR}^t, m_{\tilde{t_1}})$-plane for $\mu=-100$ and
$\tan\beta=30$. The mixing angle $\theta_t$ and $A_t$
are variable and the other fixed parameters are as in Fig.2a.
The shaded area is excluded by the triple condition
$M_{\tilde{t}_R}^2>0$, $m_{\tilde{t}_1}\geq M_Z/2$ and eq.(\ref{eq:MLR}).

\end{itemize}

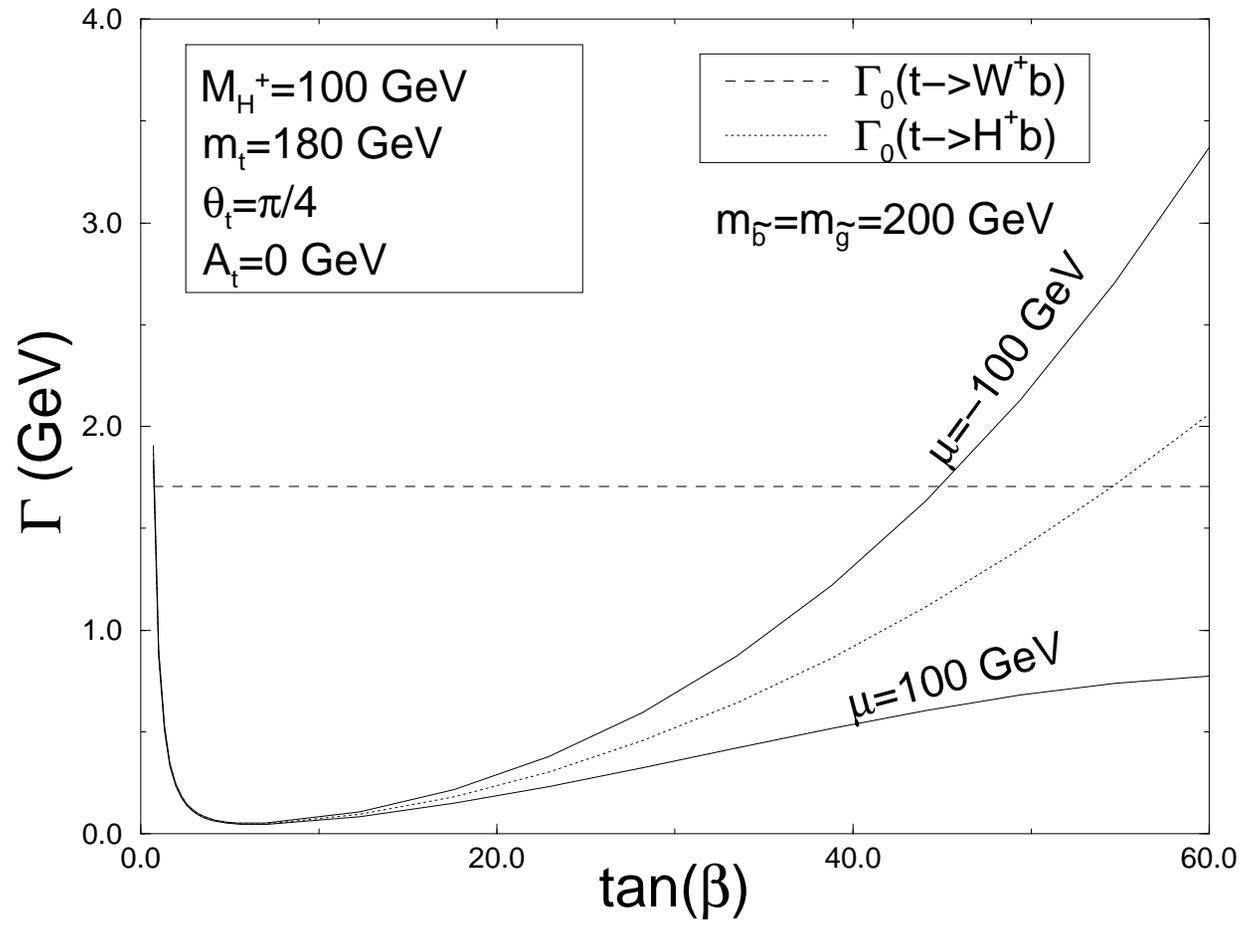
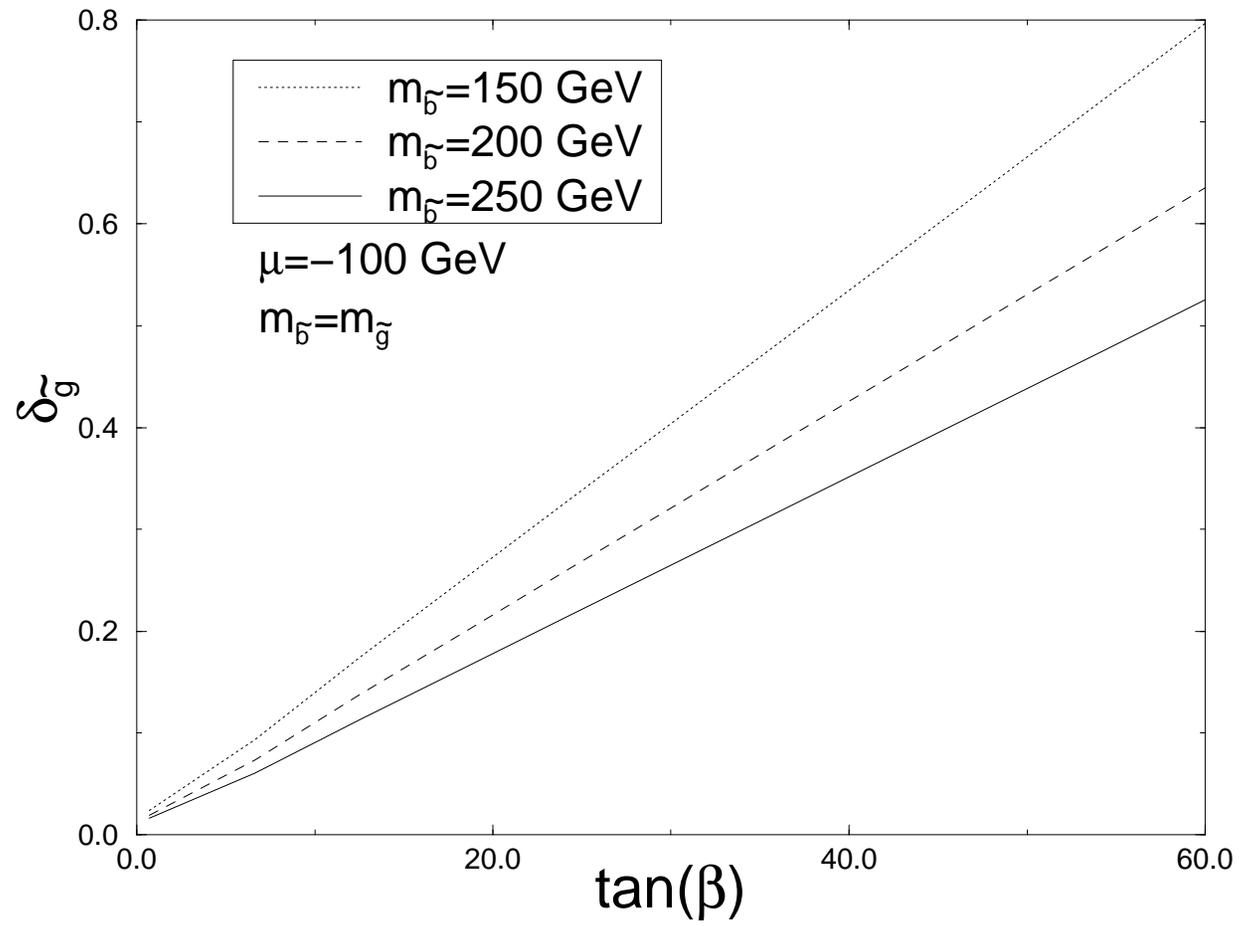
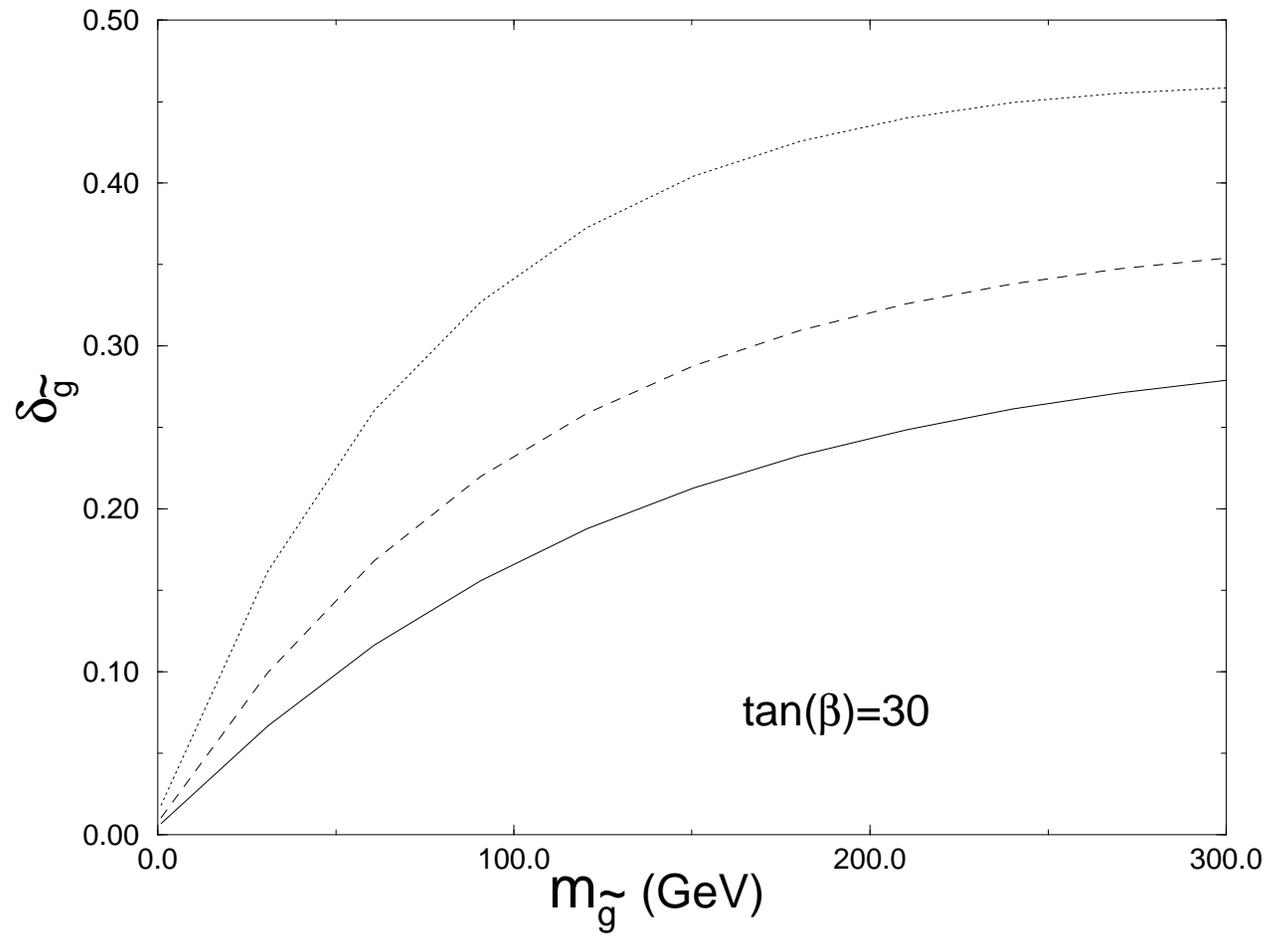
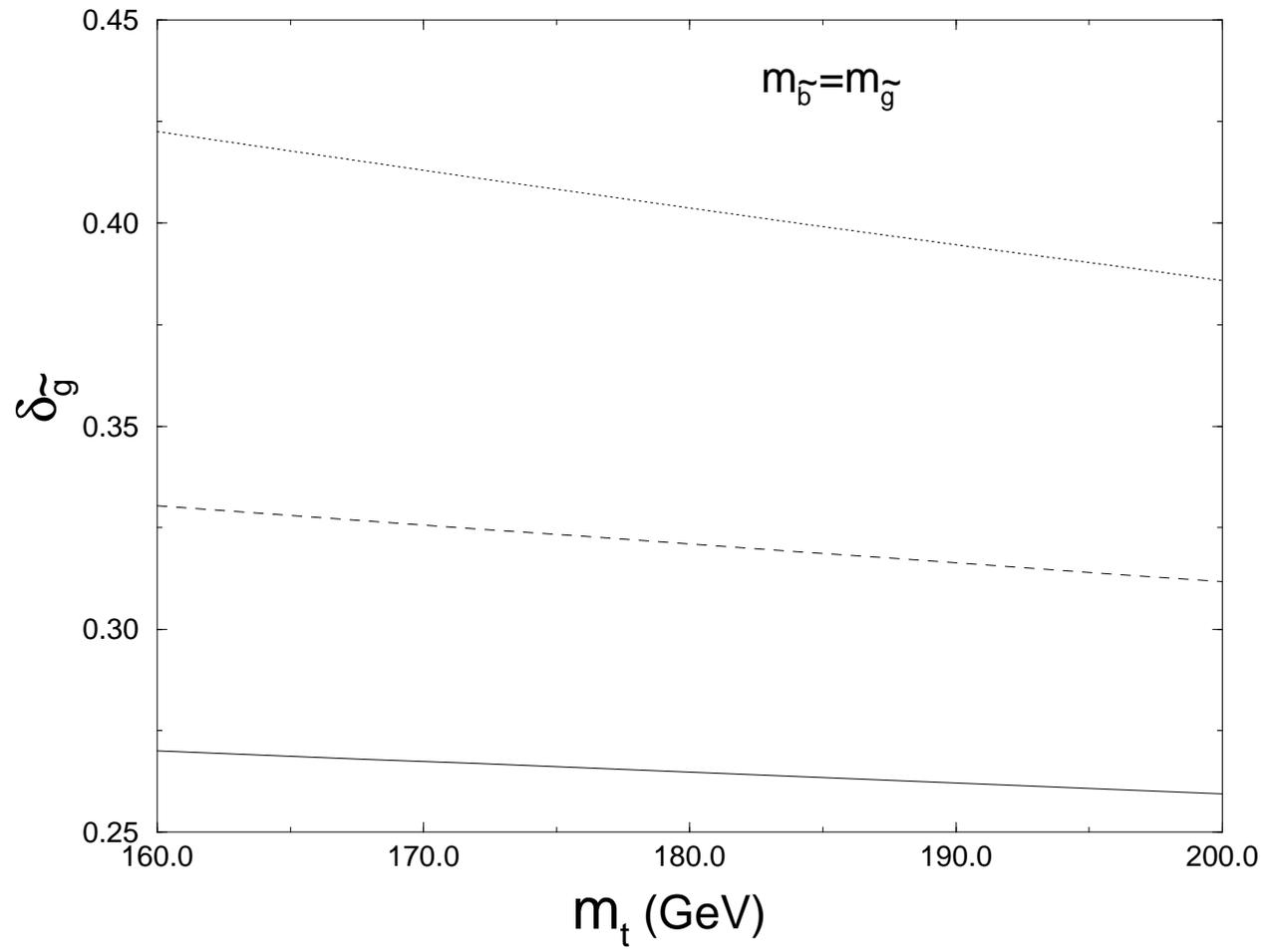
\end{document}